\documentclass[12pt]{article}
\usepackage{epsfig}
\begin{document} 
\newcommand{\ket}[1]{|#1\rangle}
\newcommand{\bra}[1]{\langle#1|}
\newcommand{\braket}[2]{\langle#1|#2\rangle}
\newcommand{\kb}[2]{|#1\rangle\langle#2|}
\newcommand{\kbs}[3]{|#1\rangle_{#3}\phantom{i}_{#3}\langle#2|}
\newcommand{\kets}[2]{|#1\rangle_{#2}}
\newcommand{\bras}[2]{\phantom{i}_{#2}\langle#1|}
\newcommand{\af}{\alpha}
\newcommand{\bt}{\beta}
\newcommand{\gm}{\gamma}
\newcommand{\la}{\lambda}
\newcommand{\dt}{\delta}
\newcommand{\s}{\sigma}
\newcommand{\qq}{(\s_{y}\otimes\s_{y})}
\newcommand{\uu}{\rho_{12}\qq\rho_{12}^{*}\qq}
\newcommand{\tr}{\textrm{Tr}} 
\centerline{\large{\bf Information transfer in leaky atom-cavity systems}}

\vskip 0.2in

\centerline{B. Ghosh\footnote{Email: biplab@bose.res.in}, A. S. 
Majumdar\footnote{Email: archan@bose.res.in} and N. 
Nayak\footnote{Email: nayak@bose.res.in}}

\vskip 0.2in

\centerline{S. N. Bose National Centre for Basic Sciences,
Salt Lake, Kolkata 700 098, India}
\date{\today}

\vskip 0.5cm                              
\begin{abstract}
We consider first a system of two enatangled cavities and a single two-level
atom
passing through one of them. A  ``monogamy'' inequality for this tripartite
system is quantitatively studied and verified in the presence of cavity
leakage. We next consider the simultaneous passage of two-level atoms
through both the cavities. Entanglement swapping is observed between the
the two-cavity and the two-atom system. Cavity dissipation leads to the
quantitative reduction of information transfer though preserving the basic 
swapping property.
\end{abstract}                                                                 

\section{Introduction}

Quantum entanglement is endowed with certain curious features.
Unlike classical correlations, 
quantum entanglement can not be freely shared among many quantum systems.
It has been observed that a quantum system being entangled with another one 
limits its possible entanglement with a third system. This has been dubbed 
the ``monogamous nature of entanglement'' which was first proposed by 
Bennett\cite{bennett}. 
If a pair of two-level quantum systems $A$ and $B$ have a perfect quantum 
correlation, namely, if they are in a maximally entangled state 
$\Psi^{-}=(\ket{01}-\ket{10})/\sqrt{2}$, then the system $A$ cannot be 
entangled 
to a third system $C$.  This indicates that there is a limitation in the 
distribution of entanglement, and several efforts have been devoted 
to capture this unique property of  ``monogamy of quantum 
entanglement'' in a quantitative way for tripartite and multipartite
systems\cite{brub,ckw,osborne}. Another distinctive property of quantum
entanglement for multipartite systems is the possibility of entanglement
swapping between two or more pairs of qubits. Using this property, two
parties that never interacted in their history can be entangled\cite{pan}.
There may indeed exist a deeper connection between the characteristics of
``monogamy'' and entanglement swapping since the features of the distribution
and transfer of quantum information is essentially reflected in the both
these properties.

Practical realization of various features of quantum entanglement are
obtained in atom-photon interactions in optical and microwave
cavities\cite{raimond}. Recently, some studies have been performed
to quantify the entanglement obtained between atoms through 
atom-photon interactions in cavities\cite{masiak,datta}. An important
attribute of real devices in the ubiquitous presence of dissipative
effects in them.  These have to be monitored in order for the effects of
quantum correlations to survive till detection. The consequences of
cavity leakage on information transfer in the micromaser has been 
quantified recently\cite{datta}. It is natural to expect  the other 
characteristics
of entanglement such as its ``monogamous'' nature, and also its exchange
or swapping to be affected by dissipative processes. Atom-photon interactions
in cavities are a sound arena for the quantitative investigations of
different aspects of quantum entanglement in realistic situations. 

With the above motivation we perform a quantitative study of the mono-gamy
of quantum entanglement and its swapping in dissipative atom-photon 
interactions in microwave cavities. We focus on a system of two entangled
single-mode cavities which are empty initially. We then consider the passage 
of a
two-level atom through either or both of them. In the next section we first
consider a tripartite pure system (two ideal cavities and one atom) and 
study the
features of ``monogamy'' exhibited between the atom-cavity and the 
cavity-cavity entanglements. In particular, we demonstrate the applicability
of the Coffman-Kundu-Wootters (CKW)\cite{ckw} ``monogamy'' inequality to this 
system. 
We next consider a realistic cavity with photon leakage, and repeat the
above analysis keeping in mind the recently conjectured validity of the
CKW inequality extended for mixed states\cite{osborne}. We find that cavity 
dissipation could lead to interesting possibilities, such as the 
enhancement of  the entanglement between the atom
and the cavity mode that it interacts with, a feature that could 
be understood by the ``monogamous'' behaviour of entanglement.
In section IV we consider a four-qubit system (two cavities and two atoms)
where our goal is to observe entanglement swapping, or the transfer of
entanglement from the initially entangled two cavities to the two atoms.
Here again, we first perform the analysis with ideal cavities, and then
consider the effects of cavity leakage on entanglement swapping.
We present some concluding remarks in section V.

\section{Monogamy of entanglement in a system of two cavities and a single atom}

\subsection{Pure state of three qubits}

We first consider two ideal cavities which can be maximally entangled
by sending a single circular Rydberg atom prepared in the exited state through
two identical and initially empty high-Q cavities 
($C_1$ and $C_2$)\cite{davidovich}.
The initial state of the two-cavity entangled system can be written as 
\begin{eqnarray}
\ket{\Psi}_{C_1C_2}=\frac{1}{\sqrt2}(\ket{0_{1}1_{2}}+\ket{1_{1}0_{2}}),  
\end{eqnarray}
where the index 1 and 2 refer to the first and second cavity, respectively.
In this set-up we consider the passage of
a two-level Rydberg atom $A_1$ prepared in the ground state
$\ket{g}$ through the cavity $C_1$. 
We are considering the resonant interaction between the two-level atom
and cavity mode frequency. The interaction Hamiltonian in the rotating frame
approximation for the atom-cavity system is 
\begin{eqnarray}
H_I=g(\s^+ a+\s^-a^\dagger),
\end{eqnarray}
where $a^\dagger$ and  $a$ are usual creation and destruction operators of the
radiation field and $\s^+(\s^-)$ are atomic operators analogous to the
Pauli spin raising and lowering operators obeying the commutation relation
$[\s^+,\s^-]=2\s_z$, where $\s_z=+1/2(-1/2)$ represents the atom in the
upper (lower) state.  $g$ is the atom-field 
interaction constant
 (or $gt$ the Rabi angle). The dynamics of the atom-photon interaction is
governed by the equation 
\begin{eqnarray}
\dot\rho = -i[H_I,\rho]
\end{eqnarray} 
with joint three-party initial ($t=0$) state corresponding to
\begin{eqnarray}
\ket{\Psi(t=0)}_{C_1C_2A_1}=\frac{1}{\sqrt2}(\ket{0_{1}1_{2}}+\ket{1_{1}0_{2}})
\otimes\ket{g_1} 
\end{eqnarray}
Hence, a  two-level atom entering the empty cavity in the upper state 
($\ket{e}$) evolves to
\begin{eqnarray}
\ket{\Psi_e(t)}&=&e^{-iH_{I}t}\ket{e,0}\nonumber\\
&=&\cos(gt)\ket{e,0}+\sin(gt)\ket{g,1}
\end{eqnarray}
at some time $t$, and similarly,
a two-level Rydberg atom entering the one photon cavity in the 
ground state evolves to 
\begin{eqnarray}
\ket{\Psi_g(t)}&=&e^{-iH_{I}t}\ket{g,1}\nonumber\\
&=&\cos(gt)\ket{g,1}-\sin(gt)\ket{e,0}
\end{eqnarray}

\vskip 0.4cm

\begin{figure}[h]
	\begin{center}
	\includegraphics[width=8cm]{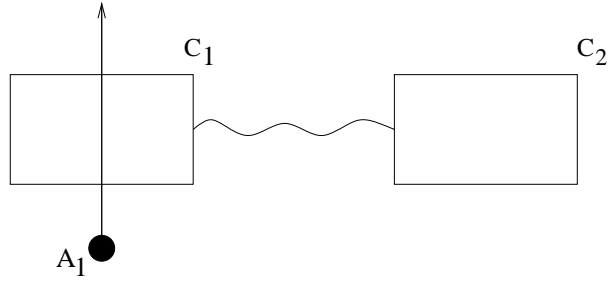}
         \caption{A two-level Rydberg atom prepared in the ground state
 is passing throuh the one of the maximally entangled cavities $C_1$}
\end{center}
\label{f1}
\end{figure}

For any interaction time $t$ the evolved state is  given by
\begin{eqnarray}
\ket{\Psi(t)}_{C_1C_2A_1}=&&\frac{1}{\sqrt2}(\ket{0_{1}1_{2}g_{1}}+
\cos{gt}\ket{1_{1}0_{2}g_{1}}\nonumber
\\
&&-\sin{gt}\ket{0_{1}0_{2}e_{1}}) 
\end{eqnarray}
\begin{eqnarray} 
\rho(t)_{C_1C_2A_1}=\kb{\Psi(t)_{C_1C_2A_1}}{\Psi(t)_{C_1C_2A_1}}
\end{eqnarray}

The reduced density states of the pairs $C_1C_2$, $C_2A_1$, $C_1A_1$ are 
given by
\begin{eqnarray} 
\rho(t)_{C_1C_2}&=&\textrm{Tr}_{A_1}(\rho(t)_{C_1C_2A_1}),\nonumber\\
&=&\frac{1}{2}\kb{0_11_2}{0_11_2}+\frac{\cos^2{gt}}{2}\kb{1_10_2}{1_10_2}
\nonumber\\
&+&\frac{\sin^2{gt}}{2}\kb{0_10_2}{0_10_2}+\frac{\cos{gt}}{2}\kb{0_11_2}{1_10_2}\nonumber\\
&+&\frac{\cos{gt}}{2}\kb{1_10_2}{0_11_2}. 
\end{eqnarray}
\begin{eqnarray}
\rho(t)_{C_2A_1}&=&\textrm{Tr}_{C_1}(\rho(t)_{C_1C_2A_1}),\nonumber\\
&=&\frac{1}{2}\kb{1_2g_1}{1_2g_1}+\frac{\cos^2{gt}}{2}\kb{0_2g_1}{0_2g_1}
\nonumber\\
&+&\frac{\sin^2{gt}}{2}\kb{0_2e_1}{0_2e_1}-\frac{\sin{gt}}{2}\kb{1_2g_1}{0_2e_1}
\nonumber\\
&-&\frac{\sin{gt}}{2}\kb{0_2e_1}{1_2g_1}.
\end{eqnarray}
\begin{eqnarray}  
\rho(t)_{C_1A_1}&=&\textrm{Tr}_{C_2}(\rho(t)_{C_1C_2A_1}),\nonumber\\
&=&\frac{1}{2}\kb{0_1g_1}{0_1g_1}+\frac{\cos^2{gt}}{2}\kb{1_1g_1}{1_1g_1}
\nonumber\\
&+&\frac{\sin^2{gt}}{2}\kb{0_1e_1}{0_1e_1}-\frac{\sin{gt}\cos{gt}}{2}\kb{1_1g_1}{0_1e_1}\nonumber\\
&-&\frac{\sin{gt}\cos{gt}}{2}\kb{0_1e_1}{1_1g_1}.
\end{eqnarray}
We now compute the mixed-state bipartite entanglement measure 
(Concurrence)\cite{hill} for different pairs. These are given by
\begin{eqnarray}
{\it C}(\rho(t)_{C_1C_2})&=&|\cos{gt}|, \\
{\it C}(\rho(t)_{C_2A_1})&=&|\sin{gt}|, \\ 
{\it C}(\rho(t)_{C_1A_1})&=&|\cos{gt}\sin{gt}|
\end{eqnarray}
and are plotted in Figure2 for varying Rabi angle, clearly reflecting
the monogamous nature of entanglement between $C_1C_2$
and $C_2A_1$.
\vskip 0.2in
\begin{figure}[t]
\begin{center}
\includegraphics[width=6cm]{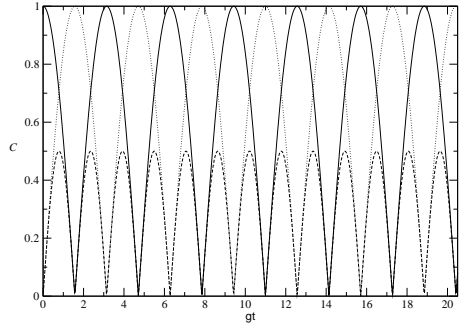}
\caption{${\it C}(\rho(t)_{C_1C_2})$ (solid line), 
${\it C}(\rho(t)_{C_2A_1})$, (dotted line), ${\it C}(\rho_{C_1A_1})$ 
(broken line) plotted with respect to the Rabi angle $gt$.}
\end{center}
\label{f2}
\end{figure}
\vskip 0.2in

The CKW inequality\cite{ckw} for the tripartite pure state 
${\rho(t)}_{C_2C_1A_1}$:\\
${{\it C}^2_{C_2C_1}}+{{\it C}^2_{C_2A_1}}\le{{\it C}^2_{C_2(C_1A_1)}}$ \\
reduces to $\cos^2{gt}+\sin^2{gt}=1$ in this case.

\subsection{Effects of cavity dissipation on entanglement}

Let us now investigate the above case in presence of the cavity 
dissipation. Since the lifetime of a two-level Rydberg atom
is usually much longer compared to the atom-cavity interaction time, we can 
safely neglect the atomic dissipation. 
The dynamics of the flight of the atom is governed by the evolution 
equation
\begin{eqnarray}
\dot\rho=\dot\rho|_{\textrm{atom-field}}+\dot\rho|_{\textrm{field-reservoir}},
\end{eqnarray} 
where the strength of the couplings are given by the parameters $\kappa$ 
(the cavity leakage constant) and $g$ (the atom-field interaction constant).
At temperature $T=0K$  the average thermal photon
number is zero, and hence one has\cite{agrawal} 	
\begin{eqnarray}
\dot\rho|_{\textrm{field-reservoir}}=-\kappa(a^\dagger a \rho-2a\rho a^\dagger+
\rho a^\dagger a).
\end{eqnarray} 
When $g \gg \kappa$, it is possible to make a secular 
approximation\cite{haroche} while solving the complete evolution 
equation by combining Eqs.(3) and (16) in order to get the density 
elements of $\rho(t)_{C_1C_2A_1}$.
We also work under a further approximation (that is justified when the cavity
is close to $0K$)  that the
probability of getting two or more photons  inside the cavities 
is zero, or in other words, a cavity always remains in the two level state 
comprising of $|0>$ and $|1>$. The tripartite (mixed) state is then obtained 
to be 
\begin{eqnarray}
\rho(t)_{C_1C_2A_1}&&=\alpha_1\kb{0_11_2g_1}{0_11_2g_1}\nonumber\\
&&+\alpha_2\kb{1_10_2g_1}{1_10_2g_1}\nonumber\\
&&+\alpha_3\kb{0_10_2e_1}{0_10_2e_1}\nonumber\\
&&+\alpha_4\kb{0_11_2g_1}{1_10_2g_1}\nonumber\\
&&+\alpha_4\kb{1_10_2g_1}{0_11_2g_1}\nonumber\\
&&+\alpha_5\kb{1_10_2g_1}{0_10_2e_1}\nonumber\\
&&-\alpha_5\kb{0_10_2e_1}{1_10_2g_1}\nonumber\\
&&+\alpha_6\kb{0_10_2e_1}{0_11_2g_1}\nonumber\\
&&-\alpha_6\kb{0_11_2g_1}{0_10_2e_1},
\end{eqnarray}
where the $\alpha_i$ are given by
\begin{eqnarray*}
\alpha_1&=&(1-\frac{e^{-\kappa_1 t}}{2})e^{-2\kappa_2 t},\\
\alpha_2&=&(\cos^2{gt})e^{-\kappa_1 t}(1-\frac{e^{-2\kappa_2 t}}{2}),\\
\alpha_3&=&(\sin^2{gt})e^{-\kappa_1 t}(1-\frac{e^{-2\kappa_2 t}}{2}),\\
\alpha_4&=&\frac{(\cos{gt})e^{-\frac{\kappa_1 t}{2}}e^{-\kappa_2 t}}{2},\\
\alpha_5&=&i(\sin{2gt})e^{-\kappa_1 t}(1-\frac{e^{-2\kappa_2 t}}{2}),\\
\alpha_6&=&i(\frac{e^{-\frac{\kappa_1 t}{2}}\sin{gt}}{2}-\frac{\kappa_1e^{-\frac{\kappa_1 t}{2}}\cos{gt}}{4g}+\frac{\kappa_1}{4g})e^{-\kappa_2 t},
\end{eqnarray*}
$\kappa_1$ and $\kappa_2$ are the leakage constants for cavity $C_1$ and 
$C_2$ respectively. 
The reduced density states of the pairs $C_1C_2$, $C_2A_1$, $C_1A_1$ are 
thus given by
\begin{eqnarray} 
\rho(t)_{C_1C_2}&=&\textrm{Tr}_{A_1}(\rho_{C_1C_2A_1}),\nonumber\\
&=&\alpha_1\kb{0_11_2}{0_11_2}+\alpha_2\kb{1_10_2}{1_10_2}
\nonumber\\
&+&\alpha_3\kb{0_10_2}{0_10_2}+\alpha_4\kb{0_11_2}{1_10_2}\nonumber\\
&+&\alpha_4\kb{1_10_2}{0_11_2}. 
\end{eqnarray}
\begin{eqnarray}
\rho(t)_{C_2A_1}&=&\textrm{Tr}_{C_1}(\rho_{C_1C_2A_1}),\nonumber\\
&=&\alpha_1\kb{1_2g_1}{1_2g_1}+\alpha_2\kb{0_2g_1}{0_2g_1}
\nonumber\\
&+&\alpha_3\kb{0_2e_1}{0_2e_1}-\alpha_6\kb{1_2g_1}{0_2e_1}
\nonumber\\
&+&\alpha_6\kb{0_2e_1}{1_2g_1}.
\end{eqnarray}
\begin{eqnarray}  
\rho(t)_{C_1A_1}&=&\textrm{Tr}_{C_2}(\rho_{C_1C_2A_1}),\nonumber\\
&=&\alpha_1\kb{0_1g_1}{0_1g_1}+\alpha_2\kb{1_1g_1}{1_1g_1}
\nonumber\\
&+&\alpha_3\kb{0_1e_1}{0_1e_1}+\alpha_5\kb{1_1g_1}{0_1e_1}\nonumber\\
&-&\alpha_5\kb{0_1e_1}{1_1g_1}.
\end{eqnarray}
and one can obtain the respective concurrences. These, namely, 
${\it C}(\rho(t)_{C_1C_2})$, ${\it C}(\rho(t)_{C_1A_1})$, and 
${\it C}(\rho(t)_{C_2A_1})$  are plotted with respect to the Rabi angle $gt$
in Figure3. As expected, dissipation reduces the respective 
concurrences. However, the ``monogamous'' character, or the `complementarity'
between ${\it C}(\rho(t)_{C_1C_2})$ and ${\it C}(\rho(t)_{C_2A_1})$ is
maintained even with cavity leakage.
\vskip 0.5cm

\begin{figure}[h]
\begin{center}
\includegraphics[width=8cm]{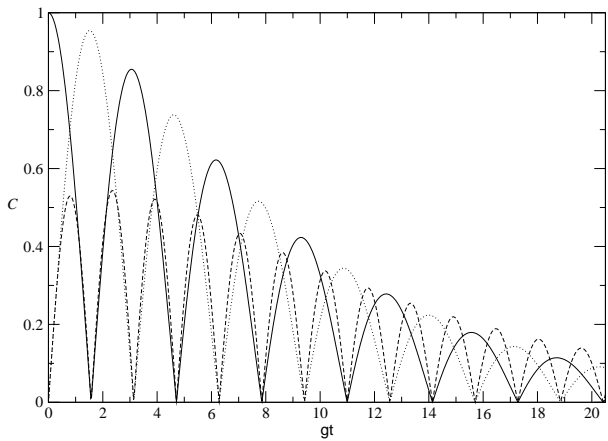}
\caption{${\it C}(\rho(t)_{C_1C_2})$ (solid line), 
${\it C}(\rho(t)_{C_2A_1})$, (dotted line), ${\it C}(\rho(t)_{C_1A_1})$ 
(broken line) plotted with respect to the Rabi angle $gt$. 
$\frac{\kappa_1}{g}=\frac{\kappa_2}{g}=0.1$.}
\end{center}
\label{f3}
\end{figure}

\vskip 0.2in

To verify the CKW inequality for the mixed state  $\rho(t)_{C_1C_2A_1}$, one 
has
to average ${\it C}(\rho(t)_{C_2(C_1A_1)})$ over all pure state 
decompositions\cite{osborne}. We however, adopt an utilitarian point of
view, and for small $\kappa$ take 
${\it C}(\rho(t)_{C_2(C_1A_1)})\approx 2\sqrt{\textrm {det}\rho_{C_2}}$. 
Note that this result holds exactly for a pure 
state\cite{ckw}. Nevertheless, for a small value of $\kappa$ and for a 
bipartite photon field,
one stays very close to a pure state.
In Figure4 we plot the left and the right  hand sides
(${{\it C}^2_{C_2C_1}}+{{\it C}^2_{C_2A_1}}$  and  
${\it C}^2_{C_2(C_1A_1)}$ respectively),
of the corresponding CKW inequality and observe that it always holds
under the above approximation.

\vskip 0.3in
\begin{figure}[h]
\begin{center}
\includegraphics[width=8cm]{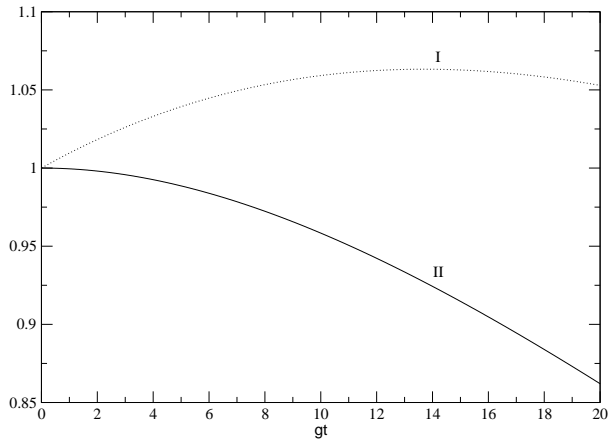}
\caption{${{\it C}^2_{C_2C_1}}+{{\it C}^2_{C_2A_1}}$ (solid line), 
${\it C}^2_{C_2(C_1A_1)})$ (dotted line) plotted with respect to 
the Rabi angle $gt$.}
\end{center}
\label{f4}
\end{figure}
\vskip 0.2in

An interesting feature of the entanglement obtained between
the atom $A1$ and the cavity $C1$ through which it interacts directly
is displayed in Figure5 where ${\it C}(\rho(t)_{A_1C_1})$
is plotted versus the dissipation parameter $\kappa$. Note that
the concurrence increases for increasing cavity loss. This happens
because the cavity leakage reduces the intial entanglement between
$C_1$ and $C_2$, and hence makes room for the subsequent entanglement 
between $C_1$ and $A_1$ to form. The dissipative mechanism is thus a striking
confirmation of the ``monogamous'' character of entanglement.
The role of the dissipative environment in creating desired forms of
entanglement has been revealed earlier in the literature\cite{braun}. 
The present
case can be also viewed as a further example of this kind.

\vskip 0.2in

\begin{figure}
\begin{center}
\includegraphics[width=7cm]{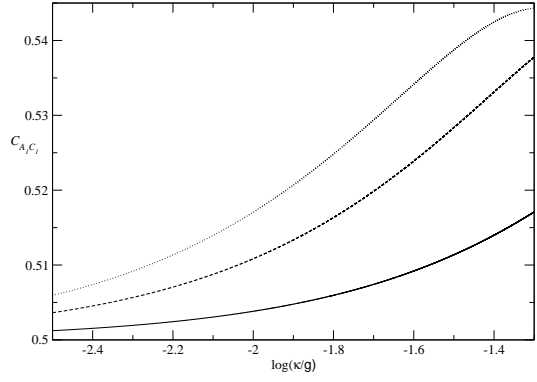}
\caption{${\it C}(\rho(t)_{A_1C_1})$ (solid line) for $gt=\pi/4$, 
${\it C}(\rho(t)_{A_1C_1})$ (broken line) for $gt=3\pi/4$,
${\it C}(\rho(t)_{A_1C_1})$, (dotted line) for $gt=5\pi/4$ 
 plotted with respect to $log(\kappa/g)$,
where $\kappa/g=\kappa_1/g=\kappa_2/g$.}
\end{center}
\label{f5}
\end{figure}
\vskip 0.2in

\section{Entanglement swapping in a system of two cavities and two atoms}

\subsection{Ideal case of four qubits}

In this section we will consider a few aspects of entanglement swapping
or the  transfer of entanglement from the two-cavity to the two-atom
system. Such a scheme can be affected by
sending two Rydberg atoms $A_1$, $A_2$  prepared in their ground 
states $g_1$, $g_2$ through two maximally entangled cavities $C_1$, $C_2$
respectively. The time of flights for the atoms  through the cavities are same.
So at $t=0$, the state of the total system is
\begin{eqnarray} 
\ket{\Psi(t=0)}_{C_1C_2A_1A_2}=\frac{1}{\sqrt2}(\ket{0_{1}1_{2}}+
\ket{1_{1}0_{2}})\otimes\ket{g_1 g_2} 
\end{eqnarray} 

\vskip 0.2in

\begin{figure}[h]
\begin{center}
\includegraphics[width=8cm]{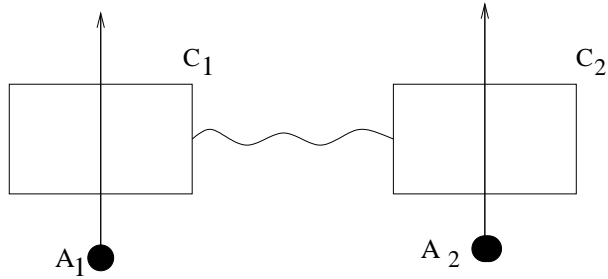}
\caption{Two Rydberg atoms $A_1$, $A_2$  prepared in the ground 
states $g_1$, $g_2$ through two maximally entangled cavities $C_1$, $C_2$
respectively.}
\end{center}
\label{f6}
\end{figure}

\vskip 0.2in

For any interaction time $t$ the evolved state is                           
\begin{eqnarray} 
\ket{\Psi(t)}_{C_1C_2A_1A_2}&&=\frac{1}{\sqrt2}(\cos{gt}\ket{0_1 1_2 g_1 g_2}
-\sin{gt}\ket{0_1 0_2 g_1 e_2}\nonumber\\
&&+\cos{gt}\ket{1_1 0_2 g_1 g_2}-\sin{gt}\ket{0_1 0_2 e_1 g_2}) 
\end{eqnarray} 
\begin{eqnarray}
\rho(t)_{C_1C_2A_1A_2}=\kb{\Psi(t)_{C_1C_2A_1A_2}}{\Psi(t)_{C_1C_2A_1A_2}}
\end{eqnarray}
The reduced density states of the pairs $C_1C_2$, $A_1A_2$ are 
given by
\begin{eqnarray} 
\rho(t)_{C_1C_2}&=&\textrm{Tr}_{A_1A_2}(\rho(t)_{C_1C_2A_1A_2}),\nonumber\\
&=&\frac{\cos^2{gt}}{2}\kb{0_11_2}{0_11_2}+
\frac{\cos^2{gt}}{2}\kb{1_10_2}{1_10_2}\nonumber\\
&+&\sin^2{gt}\kb{0_10_2}{0_10_2}+\frac{\cos^2{gt}}{2}\kb{0_11_2}{1_10_2}
\nonumber\\
&+&\frac{\cos^2{gt}}{2}\kb{1_10_2}{0_11_2}. 
\end{eqnarray}
\begin{eqnarray} 
\rho(t)_{A_1A_2}&=&\textrm{Tr}_{C_1C_2}(\rho_{C_1C_2A_1A_2}),\nonumber\\
&=&\cos^2{gt}\kb{g_1g_2}{g_1g_2}+\frac{sin^2{gt}}{2}\kb{g_1e_2}{g_1e_2}
\nonumber\\
&+&\frac{sin^2{gt}}{2}\kb{e_1g_2}{e_1g_2}+
\frac{sin^2{gt}}{2}\kb{g_1e_2}{e_1g_2}
\nonumber\\
&+&\frac{sin^2{gt}}{2}\kb{e_1g_2}{g_1e_2}. 
\end{eqnarray}
\begin{eqnarray}
{\it C}(\rho(t)_{C_1C_2})&=&\cos^2{gt}, \\
{\it C}(\rho(t)_{A_1A_2})&=&\sin^2{gt}.
\end{eqnarray}

\vskip 0.2in
\begin{figure}[h]
\begin{center}
\includegraphics[width=6cm]{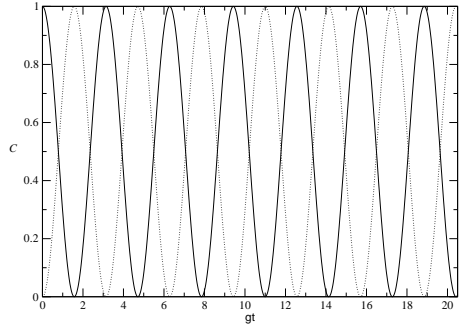}
\caption{${\it C}(\rho(t)_{C_1C_2})$ (solid line), 
${\it C}(\rho(t)_{A_1A_2})$, (dotted line) plotted with respect to 
the Rabi angle $gt$.}
\end{center}
\label{f7}
\end{figure}
\vskip 0.2in

The concurrences for the pairs $C_1$-$C_2$ and $A_1$-$A_2$ are plotted
in the Figure7.
One sees that the entanglement between two cavities are swapped by two 
atoms for the interaction times 
$gt=(2n+1)\pi/2$, ($n=0, 1, 2, \dots$).

\subsection{Information transfer with cavity dissipation}

Finally, we consider the effect of cavity leakage on the transfer
of information from the two-cavity to the two-atom system. 
Under the secular approximation\cite{haroche} and the approximation
of a two-level cavity, one can solve the master equation to obtain
the four-party density matrix which can be formally expressed as
\begin{eqnarray}
\rho(t)_{C_1C_2A_1A_2}&&=\alpha_1\kb{0_11_2g_1g_2}{0_11_2g_1g_2}\nonumber\\
&&+\alpha_2\kb{0_10_2g_1e_2}{0_10_2g_1e_2}\nonumber\\
&&+\alpha_3\kb{1_10_2g_1g_2}{1_10_2g_1g_2}\nonumber\\
&&+\alpha_4\kb{0_10_2e_1g_2}{0_10_2e_1g_2}\nonumber\\
&&+\alpha_5\kb{0_11_2g_1g_2}{1_10_2g_1g_2}\nonumber\\
&&+\alpha_5\kb{1_10_2g_1g_2}{0_11_2g_1g_2}\nonumber\\
&&+\alpha_6\kb{0_10_2g_1e_2}{0_10_2e_1g_2}\nonumber\\
&&+\alpha_6\kb{0_10_2e_1g_2}{0_10_2g_1e_2}\nonumber\\
&&+\dots
\end{eqnarray}
where the $\alpha_i$ are given by
\begin{eqnarray*}
\alpha_1&=&(1-\frac{e^{-\kappa_1 t}}{2})e^{-\kappa_2 t}\cos^2{gt},\\
\alpha_2&=&(\sin^2{gt})e^{-\kappa_2 t}(1-\frac{e^{-\kappa_1 t}}{2}),\\
\alpha_3&=&(\cos^2{gt})e^{-\kappa_1 t}(1-\frac{e^{-\kappa_2 t}}{2}),\\
\alpha_4&=&(\sin^2{gt})e^{-\kappa_1 t}(1-\frac{e^{-\kappa_2 t}}{2}),\\
\alpha_5&=&\frac{(\cos{gt})e^{-\kappa_1 t/2}e^{-\kappa_2 t/2}}{2},\\
\alpha_6&=&\frac{(e^{-\kappa_1 t/2}\sin{gt}-\frac{\kappa_1e^{-\kappa_1 t/2}}{2g}+\frac{\kappa_1}{2g})}{2}\times\\
&&(e^{-\kappa_2 t/2}\sin{gt}-\frac{\kappa_2e^{-\kappa_2 t/2}}{2g}+\frac{\kappa_2}{2g})
\end{eqnarray*}
Apart from the above eight terms no other term  contributes to either of 
the reduced 
density states $\rho_{C_1C_2}$ or $\rho_{A_1A_2}$, which are given by
\begin{eqnarray} 
\rho(t)_{C_1C_2}&=&\textrm{Tr}_{A_1A_2}(\rho(t)_{C_1C_2A_1A_2}),\nonumber\\
&=&\alpha_1\kb{0_11_2}{0_11_2}+
\alpha_3\kb{1_10_2}{1_10_2}\nonumber\\
&+&(\alpha_2+\alpha_4)\kb{0_10_2}{0_10_2}+
\alpha_5\kb{0_11_2}{1_10_2}\nonumber\\
&+&\alpha_5\kb{1_10_2}{0_11_2}. 
\end{eqnarray}
\begin{eqnarray} 
\rho(t)_{A_1A_2}&=&\textrm{Tr}_{C_1C_2}(\rho(t)_{C_1C_2A_1A_2}),\nonumber\\
&=&(\alpha_1+\alpha_3)\kb{g_1g_2}{g_1g_2}+\alpha_2\kb{g_1e_2}{g_1e_2}
\nonumber\\
&+&\alpha_4\kb{e_1g_2}{e_1g_2}+\alpha_6\kb{g_1e_2}{e_1g_2}
\nonumber\\
&+&\alpha_6\kb{e_1g_2}{g_1e_2}. 
\end{eqnarray}
\vskip 0.2in

\begin{figure}[h]
\begin{center}
\includegraphics[width=8cm]{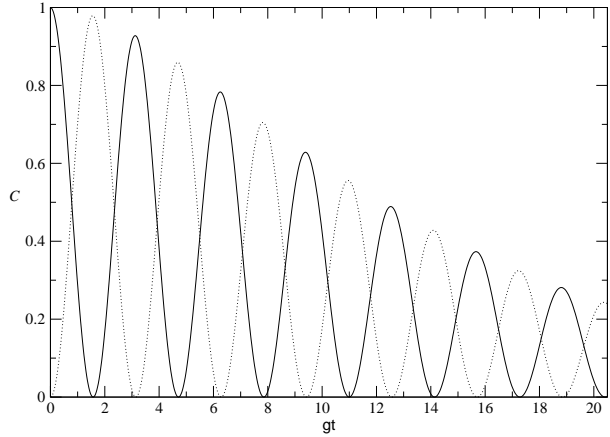}
\caption{${\it C}(\rho(t)_{C_1C_2})$ (solid line), 
${\it C}(\rho(t)_{A_1A_2})$, (dotted line) plotted with respect to 
the Rabi angle $gt$. $\kappa_1/g=\kappa_2/g = 0.1$}
\end{center}
\label{f8}
\end{figure}
\vskip 0.2in

Though the concurrences ${\it C}(\rho(t)_{C_1C_2})$ and 
${\it C}(\rho(t)_{A_1A_2})$ are reduced by the loss of cavity photons,
one sees from Figure8 that 
perfect swapping is still obtained for $gt=(2n+1)\pi/2$. One of the basic
features of information exchange between bipartite systems, represented
by entanglement swapping, is thus seen to be preserved for mixed states too.

\section{Conclusions}

In this paper we have considered two important and interesting features
of quantum entanglement, viz., ``monogamy'', and entanglement swapping.
We have used the set-up of two initialy entangled cavites\cite{davidovich} 
and a single Rydberg
atom passing through one of them to study the quantitative manifestation
of a ``monogamy'' inequality\cite{ckw} in atom-photon
interactions. The unavoidable photon leakage exists in
all real cavities used for the practical realization of quantum information
transfer. The effects of such dissipation  have been investigated
on the ``monogamous'' nature of the entanglement between the two cavities,
on one hand,
and the atom and the second cavity on the other. 
We have found that the essential 
``monogamous''
character is preserved even with cavity dissipation. We have further
seen that the entanglement between the atom and the cavity through which
it passes increases with larger dissipation, a feature that could 
be understood by invoking the ``monogamous'' character of entanglement.
We have then considered a set-up involving two entangled cavities, and
two Rydberg atoms. Entanglement swapping from the two cavities by the two 
atoms which never
interact directly with each other is observed in this system. Cavity 
dissipation, of course reduces the total amount of information exchange,
similar to the results obtained in the context of the single-atom
micromaser\cite{datta}. Moreover, here we have verified that the property
of swapping is preserved with dissipation. Further studies on different
quantitative manifestations of information transfer in the presence of
dissipative effects might be useful for the construction of realistic 
devices implementing various protocols. Practical realization of
two-cavity entanglement is in progress at the Ecole Normale 
Superieure\cite{haroche2}.


\begin{thebibliography}{99}
\bibitem{bennett}
Bennett C. H., Lecture course in the School on Quantum Physics 
and Information Processing, TIFR, Mumbai, 2002 (http://qpip-server.tcs.tifr.
res.in/ qpip/HTML/Courses/Bennett/TIFR2.pdf).
\bibitem{brub} 
V. Buzek, V. Vedral, M. B. Plenio, P. L. Knight and M. Hillery, Phys. Rev. 
A{\bf 55}, 3327 (1997);
D. Bru$\beta$, Phys. Rev. A {\bf 60}, 4344 (1999); W. D\" ur, 
G. Vidal and J. I. Cirac, Phys. Rev. A 62, 062314 (2000); 
M. Koashi, V. Buzk and N. Imoto, Phys. Rev. A {\bf 62}, 050302 (2000); 
K. A. Dennison and W. K. Wootters, Phys. Rev. A {\bf 65}, 010301R (2001);
B. M. Terhal, quant-ph/0307120; M. Koashi and A. Winter, Phys. Rev. A{\bf 69},
022309 (2004).
\bibitem{ckw}
V. Coffman, J. Kundu and W. K. Wootters, Phys. Rev. A {\bf 61}, 
052306 (2000).
\bibitem{osborne}
T. J. Osborne and F. Verstraete, e-print quant-ph/0502176 (2005).
\bibitem{pan}
J. W. Pan, D. Bouwmeester, H. Weinfurter, and A. Zeilinger, 
Phys. Rev. Lett. {\bf 80}, 3891 (1998);
E. S. Guerra and C. R. Carvalho, quant-ph/0501078.
\bibitem{raimond}
J. M. Raimond, M. Brune, and S. Haroche, Rev. Mod. Phys. {\bf 73}, 565 (2001).
\bibitem{masiak}
P. Masiak, Phys. Rev. A{\bf 66}, 023804 (2002).
\bibitem{datta}
A. Datta, B. Ghosh, A. S. Majumdar and N. Nayak, Europhys. Lett. {\bf 67}, 934
(2004).
\bibitem{davidovich}
L. Davidovich, N. Zagury, M. Brune, J.M. Raimond, and S. Haroche, Phys. Rev. A
{\bf 50}, R895 (1994); V. Giovannetti, D. Vitali, P. Tombesi, and A. Ekert,
Phys. Rev. A{\bf 62}, 032306 (2000); A. Rauschenbeutel, P. Bertet, S. Osnaghi, 
 G. Nogues, M. Brune, J.M. Raimond, and S. Haroche, Phys. Rev. A{\bf 64}, 
 050301 (2001).
\bibitem{hill}
S. Hill and W. K. Wootters, Phys. Rev. Lett. {\bf78}, 5022, (1997); 
W. K. Wootters, Phys. Rev. Lett. {\bf80}, 2245 (1998).
\bibitem{agrawal}
See, for instance, G. S. Agrawal, in {\it Springer Tracts in Modern Physics}, 
Vol.70, (Springer-Verlag, Berlin \& New York, 1974). 
\bibitem{haroche}
S. Haroche and J. M. Raimond, in ``Advances in atomic and molecular physics'',
Vol. 20, eds. D. R. Bates and B. Bederson (Academic, New York, 1985).
\bibitem{braun}
A. Beige, D. Braun, B. Tregenna and P. L. Knight, Phys. Rev. Lett. {\bf 85},
1762 (2000); D. Braun, Phys. Rev. Lett. {\bf 89}, 277901 (2002).
\bibitem{haroche2}
S. Haroche and J. M. Raimond, private communication.
\end{thebibliography}
\end{document}